\documentclass[letterpaper, twocolumn, prl, superscriptaddress]{revtex4-2}
\usepackage{graphicx, amsmath, amssymb, bm, titlesec,mathtools}
\usepackage{graphicx,amsmath, amssymb, bm, epstopdf,dsfont}

\usepackage{xcolor}
\usepackage{gensymb}
\usepackage{braket}

\begin{document}

\title{Propagation-induced entanglement revival}

\author{Abhinandan Bhattacharjee$^{1}$, Mritunjay K. Joshi$^{1}$, Suman Karan$^{1}$, Jonathan Leach$^{2}$, and Anand K. Jha$^{}$}

\email{akjha9@gmail.com}

\affiliation{$^{}$ Department of Physics, Indian Institute of Technology Kanpur, Kanpur, UP 208016, India\\
$^{2}$School of Engineering and Physical Sciences, Heriot-Watt University, Edinburgh, EH14 4AS, UK}

\date{\today}

\begin{abstract}

The practical implementation of free-space quantum information tasks requires entanglement to be sustained over long distances and in the presence of turbulent and noisy environments. The transverse position-momentum entanglement of photon pairs produced by parametric down-conversion has found several uses in quantum information science, however, it is not suitable for applications involving long-distance propagation as the entanglement decays very rapidly when photons propagate away from their source.  Entanglement is lost after a few centimetres of propagation, and the effect becomes even more pronounced in turbulent environments. In contrast, in this article, we show that entanglement in the angle-orbital angular momentum (OAM) bases exhibits a remarkably different behaviour.  As with the position-momentum case, initially, the angle-OAM entanglement decays with propagation, but as the photons continue to travel further from the source, the photons regain their strongly correlated behaviour, and the entanglement returns.   We theoretically and experimentally demonstrate this behaviour and show that entanglement returns even in the presence of strong turbulence.  The only effect of turbulence is to increase the propagation distance for revival, but once revived, the two photons remain entangled up to an arbitrary propagation distance.  This work highlights the role that OAM-angle entanglement will play in applications where quantum information is shared over long distances.

\end{abstract}

\maketitle

Quantum Entanglement \cite{einstein1935pr,horodecki2009rpm,reid2009rmp,friis2019nat} is the key resource behind the advancement of many applications such as quantum imaging \cite{moreau2019natrevphy}, quantum communication \cite{gisin2007natphot}, quantum information processing \cite{slussarenko2019apr}, and quantum computing \cite{steane199rpp}. Spontaneous parametric down-conversion (SPDC) is one of most widely used methods for generating entangled photons, in which a pump photon at a higher frequency interacts with a nonlinear crystal and produces two separate photons at lower frequencies called the signal and idler photons. The entanglement of down-converted photons has been extensively studied in the discrete  finite-dimensional bases such as polarization \cite{giustina2015prl}, time-bin \cite{thew2004prl,vedovato2018prl}, and orbital angular momentum (OAM) \cite{dada2011natphys,fickler2012science} as well as in the continuous-variable bases such as position-momentum \cite{howell2004prl,edgar2012natcom,moreau2014prl}, angle-OAM \cite{leach2010science}, radial position-radial momentum \cite{chen2019prl}, time-energy \cite{khan2006prl, maclean2018prl, mei2020prl}. Although there are several ways of quantifying two-photon entanglement in two-dimensional bases \cite{wootters1998prl}, there is no quantifier for more than two-dimensional bases and continuous-variable bases, in which cases one can talk only in terms of entanglement certifiers \cite{friis2019natrevphys}. For continuous-variable bases, there are several entanglement certifiers such as Einstein Podolsky Rosen (EPR) criterion \cite{einstein1935pr, howell2004prl, edgar2012natcom, moreau2014prl, leach2010science, bowen2003prl, fadel2020pra},  partial transpose \cite{simon2000prl, giedke2001prl}, and R{\'e}nyi entropy \cite{saboia2011pra,rastegin2017pra}. Among these certifiers, the EPR-criterion is the most widely used one, and is used even beyond photonic quantum systems  \cite{ockeloen2018nature, fadel2018science}.

The practical implementation of quantum information tasks  requires entanglement to be sustained over long distances and in  turbulent environments. The feasibility of utilizing entanglement in the finite-dimensional bases for long-distance quantum-information applications has been demonstrated in several experimental works \cite{aspelmeyer2003science, steinlechner2017natcomm,valencia2020natphot,jin2019optexp, ecker2019prx}. However, the suitability of entanglement in the continuous-variable bases for long-distance applications has not been established so far. Among the continuous-variable bases, position-momentum bases have been extensively investigated for its applicability in several applications such as quantum imaging \cite{moreau2018optexp, aspden2013njp, defienne2021pixel, toninelli2019optica, bornman2019njpqi}, quantum holography \cite{defienne2021natphys, devaux2019pra}, quantum metrology \cite{brida2010natphot}, and quantum secure communication \cite{zhang2008prl, almeida2005experimental}. Although position-momentum entanglement has found uses in many of these applications, it has not been found suitable for applications involving long-distance propagation. This is because of the fact that as the photons propagate away from the down-conversion crystal, the position-momentum entanglement decays very rapidly  \cite{walborn2010phyrep, schneeloch2016jop} and this effect becomes worse in the presence of turbulent environments.

\begin{figure*}[t!]
\includegraphics[scale=1]{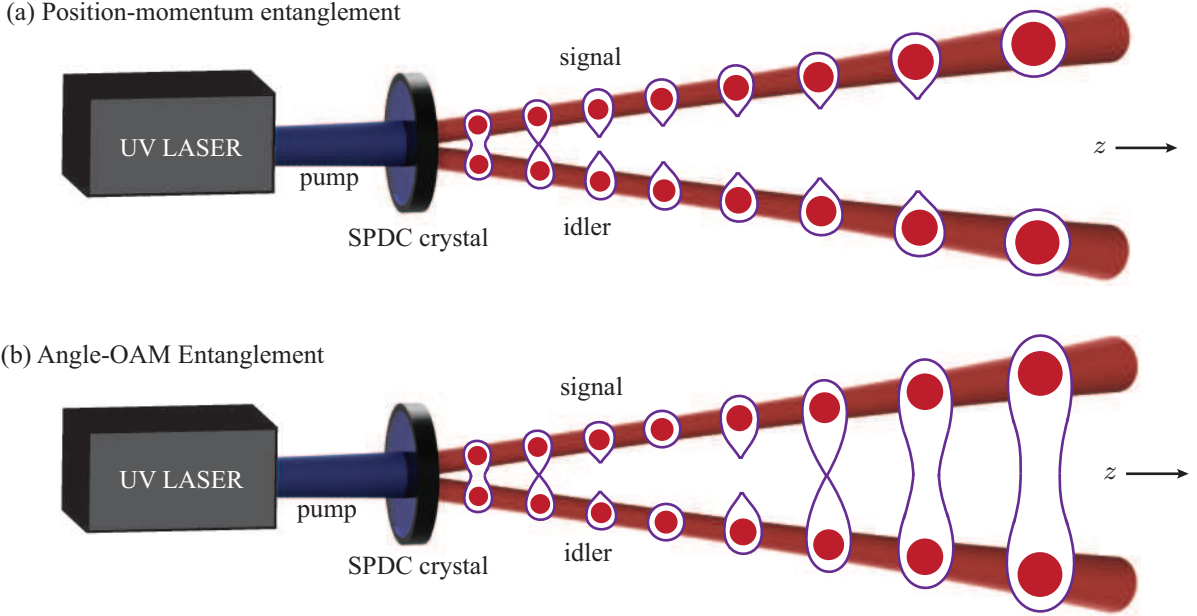}
\caption{ (a) and (b) illustrate how entanglement in the position-momentum and angle-OAM bases change as the two photons propagate away from the down-conversion crystal.}\label{FIG1}
\end{figure*}

In this article, we explore propagation of entanglement in the angle-OAM bases and demonstrate that the entanglement of down-converted photons in the angle-OAM bases exhibits a remarkably different behaviour than in the position-momentum basis. Just as in the case of position-momentum basis, initially, the angle-OAM entanglement decays with propagation, but as the photons continue to travel further from the source, the entanglement gets revived. We refer to this behaviour as the propagation-induced entanglement revival. We theoretically and experimentally demonstrate this behaviour and show that the propagation-induced entanglement revival takes place even in the presence of strong turbulence. This feature of angle-OAM entanglement can therefore have important implications for long-distance quantum information applications. Figures~\ref{FIG1}(a) and ~\ref{FIG1}(b) illustrate how entanglement in the position-momentum and angle-OAM bases propagate away from the crystal. The entanglement in the position-momentum bases decays as the photons propagate away from the crystal, and once the entanglement is lost, it does not revive upon further propagation. On the other hand, just as with the position-momentum case, the entanglement in the angle-OAM bases also decays initially as the photons start to propagate away from the crystal; however, a further propagation by some distance revives the entanglement. Once revived, the entanglement remains intact upto an arbitrary propagation distance.

\begin{figure*}[t!]
\includegraphics[scale=1]{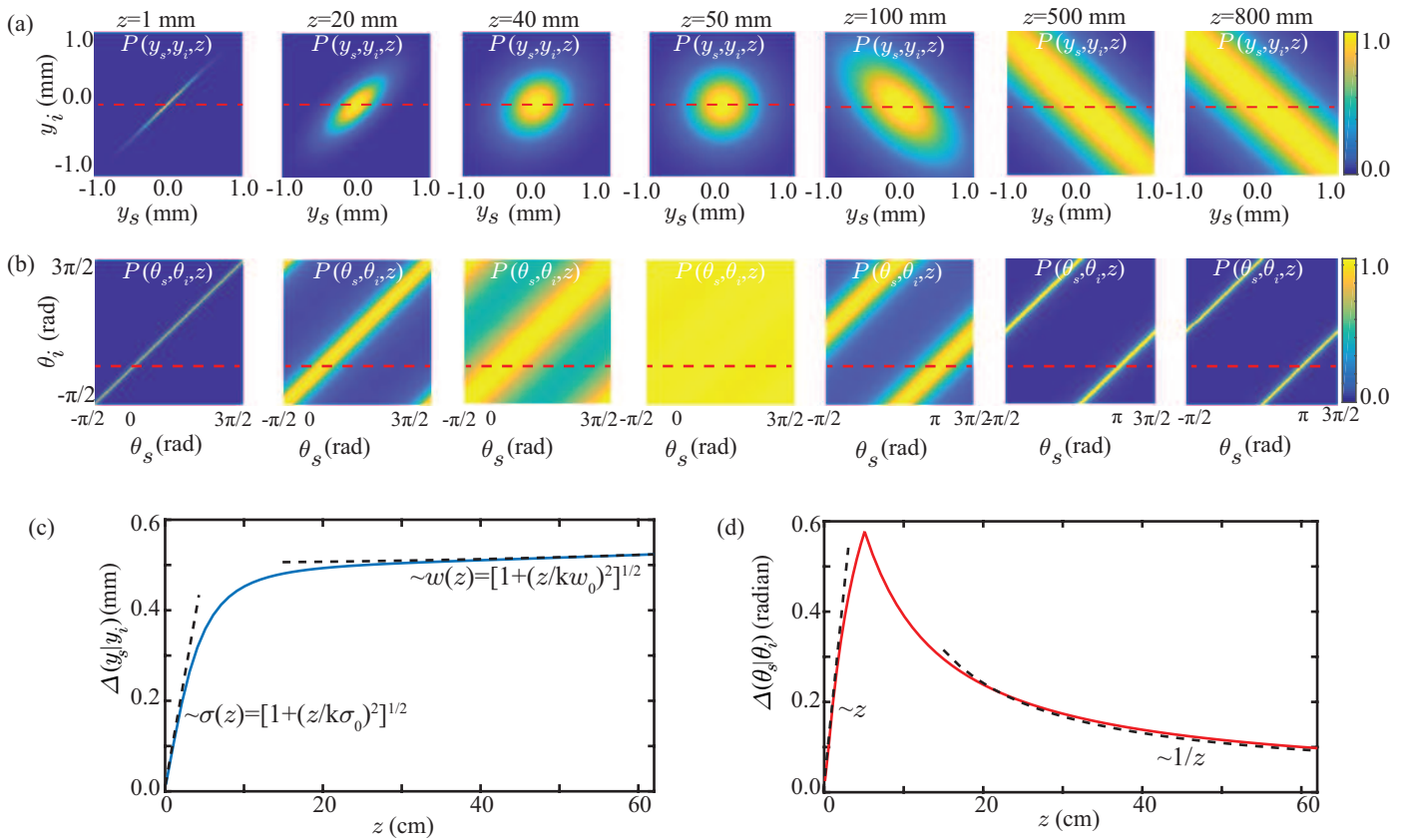}
\caption{ (a) and (b) show the two-photon position probability distribution function $P(y_s,y_i;z)$ and the angle probability distribution function $P(\theta_s,\theta_i;z)$ respectively at various $z$ values. (c) Numerically calculated conditional position uncertainty $\Delta(y_s|y_i; z)$ as a function of $z$. The two dotted lines show the $z$-scaling of the uncertainty in the near- and far-field regions. (d) Numerically calculated conditional angle uncertainty $\Delta(\theta_s|\theta_i; z)$ as a function of $z$. The two dotted lines show the $z$-scaling of the uncertainty in the near- and far-field regions.}\label{correlation}
\end{figure*}

\begin{figure*}[t!]
\includegraphics[scale=1]{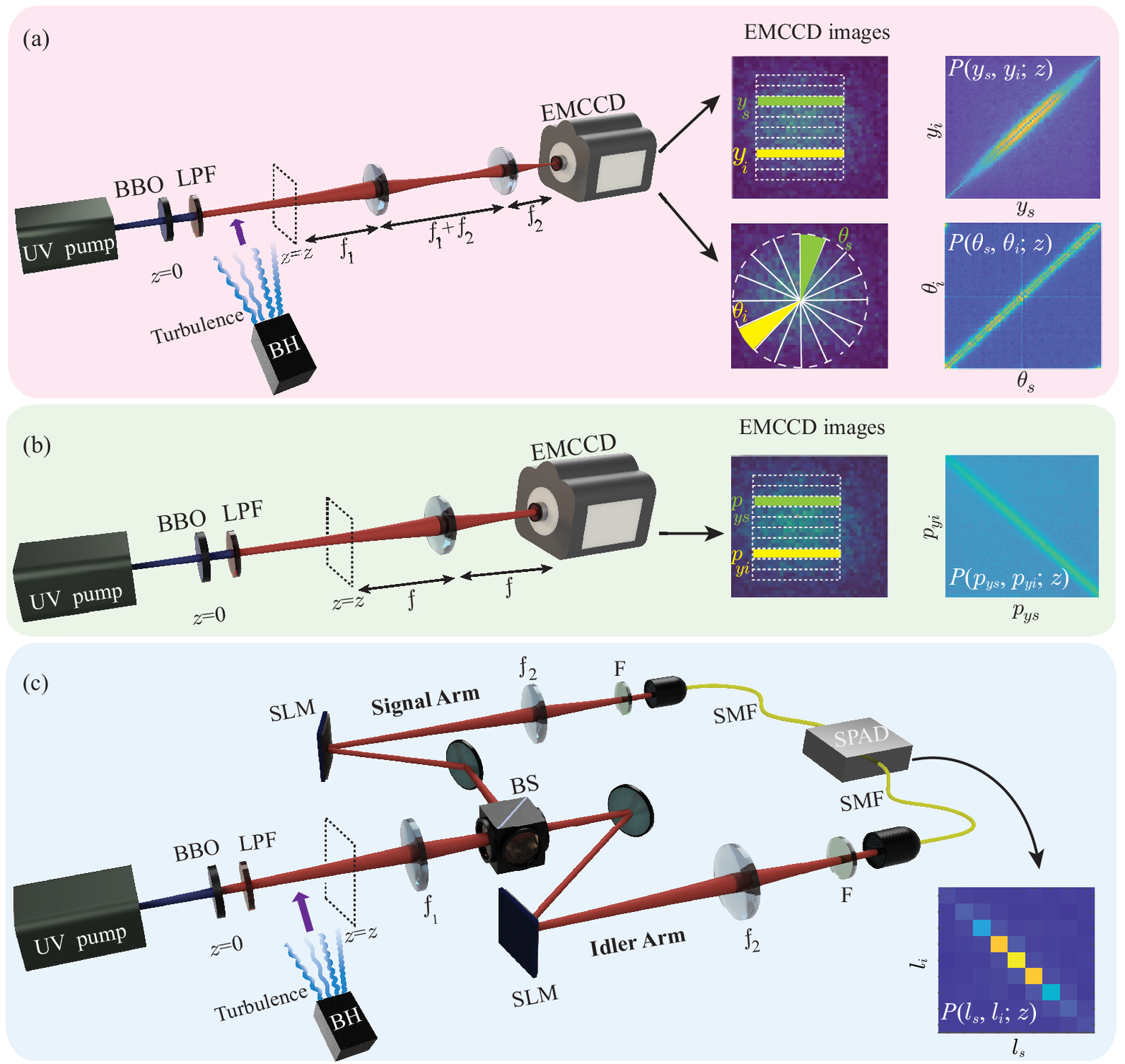}
\caption{ (a) Schematic of the experimental setup for measuring position and angle coincidences. Inset shows the EMCCD images of the SPDC field and the corresponding two-photon position and angle probability distribution functions. (b) Schematic of the experimental setup for measuring the two-photon momentum probability distribution function. Inset shows EMCCD images of the SPDC field and the corresponding two-photon momentum probability distribution function. (c) Schematic of the experimental setup for measuring OAM coincidence and the OAM correlation. LPF: long-pass filter, BS: beam splitter, SLM: spatial light modulator, SMF: single-mode fiber, F: interference filter. The blower heater (BH) is used for generating turbulence, and it is switched on in the path of the SPDC field when studying the effect of turbulence on entanglement propagation.}\label{fig_setup1}
\end{figure*} 

We now present our quantitative analysis and observations of propagation-induced entanglement revival using the two-photon field produced by SPDC. For a Gaussian pump with beam waist at the crystal plane $z=0$, the two-photon wavefunction in the position basis at a propagation distance $z$ is given by \cite{edgar2012natcom,walborn2010phyrep,schneeloch2016jop} $\psi(\bm{\rho}_s, \bm{\rho}_i;z) = A \exp\left[-\frac{|\bm{\rho}_s+\bm{\rho}_i|^2}{4 w(z)^2}\right]\exp\left[-\frac{|\bm{\rho}_s-\bm{\rho}_i|^2}{4\sigma(z)^2}\right]e^{i\phi(\bm{\rho}_s, \bm{\rho}_i, z)}$, where $\bm{\rho}_s \equiv (x_s,y_s)$ and $\bm{\rho}_i \equiv (x_i,y_i)$ are the transverse positions of the signal and idler photons, respectively at $z$, and where $w(z)=w_0\sqrt{1+z^2/(k^2 w_0^4)}$, $\sigma(z)=\sigma_0\sqrt{1+z^2/(k^2\sigma_0^4)}$, and $k=\pi/\lambda_p$. Also, $w_0$ is the pump beam waist at $z=0$, $\sigma_0=\sqrt{{0.455 L \lambda_p}/{2\pi}}$, $L$ is the length of the crystal,  $\lambda_p$ is the wavelength of the pump field, and $e^{i\phi(\bm{\rho}_s, \bm{\rho}_i, z)}$ is a phase factor. The two-photon position probability distribution function $P(\bm{\rho}_s,\bm{\rho}_i;z)=|\psi(\bm{\rho}_s, \bm{\rho}_i;z)^* \psi(\bm{\rho}_s, \bm{\rho}_i;z)|$ at  $z$ is therefore given by 
\begin{align}
P(\bm{\rho}_s,\bm{\rho}_i;z)=|A|^2 \exp\left[-\frac{|\bm{\rho}_s+\bm{\rho}_i|^2}{2w(z)^2}\right] \exp\left[-\frac{|\bm{\rho}_s-\bm{\rho}_i|^2} {2\sigma(z)^2}\right] \label{eqn_spatial_PD}
\end{align}
For a fixed idler position, say in the $y$-direction, the two-photon position probability distribution is referred to as the conditional position probability distribution of the signal photon at $z$ and is denoted as $P(y_s|y_i; z)$. The standard deviations of $P(y_s|y_i; z)$ is referred to as the conditional position uncertainty $\Delta(y_s|y_i; z)$ of the signal photon. Similarly, by writing the two-photon wavefunction $\psi(\bm{\rho}_s, \bm{\rho}_i;z)$ in the transverse momentum basis, one can calculate conditional momentum uncertainty $\Delta(p_{ys}|p_{yi}; z)$ of the signal photon at $z$. According to the EPR criterion of entanglement, if the product $\Delta(y_s|y_i; z)\Delta(p_{ys}|p_{yi}; z) < 0.5\hbar$, then the two photons are entangled in the position-momentum bases \cite{einstein1935pr, howell2004prl}. In addition to the position-momentum bases the down-converted photons are rendered entangled in the angle-OAM bases as well. Using Eq.~(\ref{eqn_spatial_PD}) and the transformations $\bm{\rho}_s=(r_s\cos\theta_s,r_s\sin\theta_s)$ and $\bm{\rho}_i=(r_i\cos\theta_i,r_i\sin\theta_i)$, we can write the two-photon angle probability distribution function $P(\theta_s,\theta_i; z)$ as:
\begin{equation}
P(\theta_s,\theta_i;z)=\iint r_s r_i P(r_s,\theta_s,r_i,\theta_i;z) dr_s dr_i,
\label{eqn_angular_PD}
\end{equation}
where $(r_s,\theta_s)$ and $(r_i,\theta_i)$ are the polar coordinates of the signal and idler photons. Using $P(\theta_s,\theta_i; z)$, one can obtain the conditional angle probability distribution function $P(\theta_s|\theta_i; z)$  and thereby the conditional angle uncertainty $\Delta(\theta_s|\theta_i; z)$ of the signal photon. Denoting the conditional OAM uncertainty by $\Delta(l_s|l_i; z)$, we write the EPR criterion for entanglement in the angle-OAM bases as: $\Delta(\theta_s|\theta_i; z)\Delta(l_s|l_i; z) < 0.5 \hbar$ \citep{leach2010science}.

Now, using Eqs.~(\ref{eqn_spatial_PD}) and ~(\ref{eqn_angular_PD}), and the relevant experimental parameters $w_0=507$ $\mu$m, $L=5$ mm and $\lambda_p=355$ nm, we numerically evaluate $P(y_s,y_i;z)$ and  $P(\theta_s,\theta_i;z)$ at different propagation distances $z$ and plot them in Figs.~\ref{correlation}(a) and ~\ref{correlation}(b) respectively. In plotting $P(y_s,y_i;z)$ and $P(\theta_s,\theta_i;z)$, we scale them in order to make their maximum values equal to one. Next, by fixing $y_i=0$ mm in $P(y_s,y_i; z)$ and $\theta_i=0$ in $P(\theta_s,\theta_i; z)$, we calculate $P(y_s|y_i;z)$ and $P(\theta_s|\theta_i;z)$ and thereby the conditional position uncertainty $\Delta(y_s|y_i; z)$ and the conditional angle uncertainty $\Delta(\theta_s|\theta_i; z)$, and plot them in Figs.~\ref{correlation}(c) and ~\ref{correlation}(d) respectively. From the plots in Figs.~\ref{correlation}(c) and ~\ref{correlation}(d), we find that as the down-converted photons propagate away from the crystal, the conditional position uncertainty increases monotonically. However, the conditional angle uncertainty increases initially but later begins to decrease monotonically. (See Supplementary Information Sec.~I and II for more detailed analysis and numerical simulations.)

 Although it is very difficult to derive the general analytical expressions for the  conditional position and angle uncertainties as a function of $z$, we derive expressions for how the conditional uncertainties scale with $z$ in the near- and far-field regions. The two dotted lines in Figs.~\ref{correlation}(c) show how the conditional position uncertainty $\Delta(y_s|y_i; z)$ scales with $z$ in the near and far fields. We find that $\Delta(y_s|y_i; z)$ increases monotonically as a function of $z$ in both the near- and far-field regions. While the uncertainty increases as $\sigma(z)$ in the near-field, it  increases as $w(z)$ in the far-field.   The two dotted lines in Figs.~\ref{correlation}(d) show how the conditional angle uncertainty $\Delta(\theta_s|\theta_i; z)$ scales with $z$ in the near and far field. We find that while $\Delta(\theta_s|\theta_i; z)$ increases as $z$ in the near field regions, it decreases as $1/z$ in the far-field regions. (For detailed theoretical calculations of the scaling laws, see Supplementary Information Sec.~I-B and I-C.)

 \begin{figure*}[t!]
\includegraphics[scale=1]{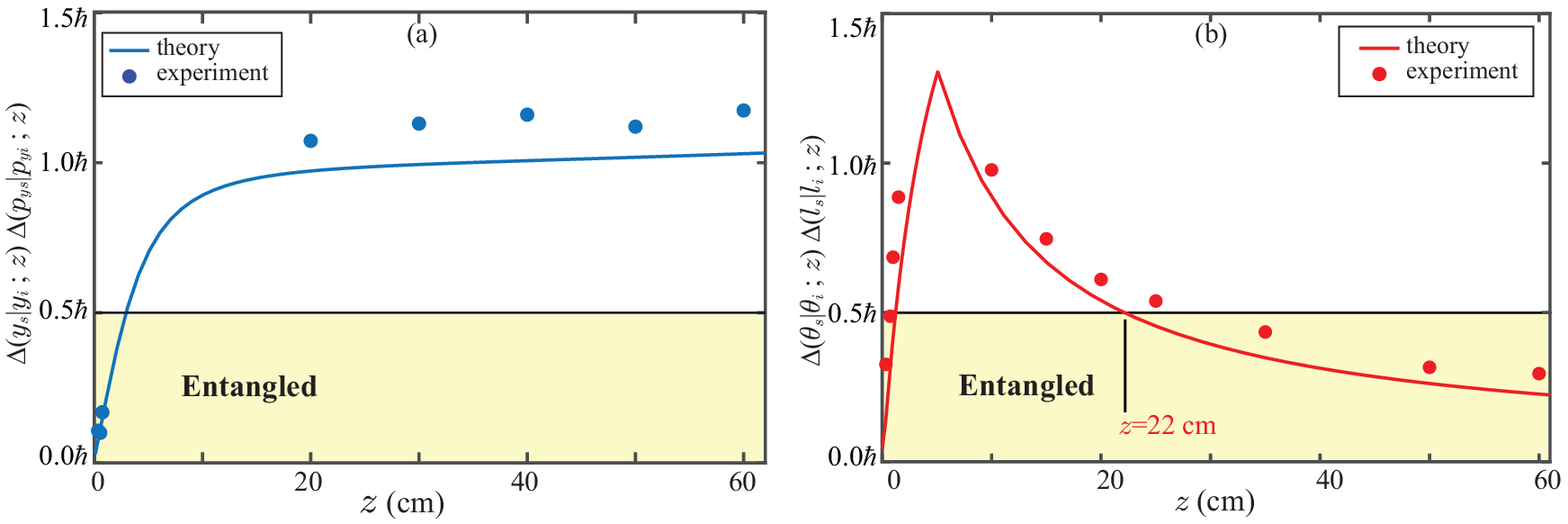}
\caption{(a) Conditional position-momentum uncertainty product $\Delta(y_s|y_i; z)\Delta(p_{sy}|p_{iy}; z)$ as a function of the propagation distance $z$. The solid dots are the experimental results and the solid line is the best theoretical fit. (b) conditional angle-OAM uncertainty product $\Delta(\theta_s|\theta_i; z)\Delta(l_{s}|l_{i}; z)$ as a function of the propagation distance $z$. The solid dots are the experimental results and the solid line is the best theoretical fit. As indicated on the plot, the theoretical prediction for entanglement revival is at $z=22$ cm while we observe it at about $z=28$ cm.}\label{FIG3}
\end{figure*}

Figure~\ref{fig_setup1} shows the schematic of the experimental setup for measuring the two-photon probability distribution functions $P(y_s,y_i;z)$, $P(p_{ys},p_{yi};z)$, $P(\theta_s,\theta_i;z)$, and $P(l_s,l_i;z)$ through coincidence measurements of the two  photons. An ultraviolet (UV) continuous wave (CW) Gaussian pump (Coherent Genesis STM UV laser) of wavelength $\lambda_p=355$ nm, beam waist $w_0=507 \ \mu$m  is incident on a $5$ mm $\times$ $5$ mm $\times$ $5$ mm $\beta$-barium borate (BBO) crystal. The crystal is cut in a manner that it produces signal and idler photons with collinear type-I phase-matching condition. A long-pass filter (LPF) is placed after the crystal to block the UV pump. We use an Andor iXon Ultra-897 electron-multiplied charged coupled device (EMCCD) camera that has a $512 \times 512$ pixel grid with each pixel being $16 \times 16$ $\mu$m$^2$ of size. A $10$ nm bandpass filter centered at $710$ nm is used in order to detect the down-converted photons. The blower heater (BH) produces turbulence by blowing hot air, and it is switched on during our experiments involving turbulence.

For the coincidence measurements of $P(y_s,y_i;z)$, $P(p_{ys},p_{yi};z)$, and $P(\theta_s,\theta_i;z)$, we use an EMCCD camera \cite{reichert2018scirpt, defienne2018prl}, as depicted in Figs.~\ref{fig_setup1}(a) and \ref{fig_setup1}(b). For measuring $P(y_s,y_i;z)$ and $P(\theta_s,\theta_i;z)$ and thereby the corresponding uncertainties $\Delta(y_s|y_i; z)$ and $\Delta(\theta_s|\theta_i; z)$, we image the transverse plane at $z$ onto the EMCCD camera plane using a $4f$-imaging system, as depicted in Fig.~\ref{fig_setup1}(a). For measuring $P(p_{ys},p_{yi};z)$ at $z$, we use a $2f$ imaging system and keep the EMCCD camera plane at the Fourier plane of the transverse plane at $z$, as depicted in Fig.~\ref{fig_setup1}(b). We then measure the two-photon position probability distribution function at the EMCCD camera plane, which is proportional to the two-photon momentum probability distribution function $P(p_{ys},p_{yi};z)$ at $z$. The conditional momentum uncertainty $\Delta(p_{ys}|p_{yi}; z)$ is obtained by multiplying the conditional position uncertainty at the EMCCD plane by $k\hbar/f$, where $f$ is the focal length of the lens. (For details regarding measurement techniques and results, see Supplementary Information Secs.~III, IV, and V). For the coincidence measurements of the two-photon OAM probability distribution $P(l_s,l_i;z)$, we make use of two electronically gated single photon avalanche diode (SPAD) detectors \cite{leach2010science, leach2012pra}, as depicted in Fig.~\ref{fig_setup1}(c). We image the transverse plane at $z$ onto the SLMs kept in the signal and idler arms. Specific holograms are displayed onto both the SLMs and then the signal and idler SLM planes are imaged onto the input facets of single-mode fibers (SMFs) kept in the signal and idler arms. The combination of the hologram and SMF in each arm projects the input field into a particular OAM mode which then gets detected by the SPAD detector through the SMF. An electronic coincidence circuit then yields the coincidence counts. By displaying different holograms on the SLMs, we measure the two-photon OAM probability distribution.

\begin{figure}[b!]
\includegraphics[scale=1.0]{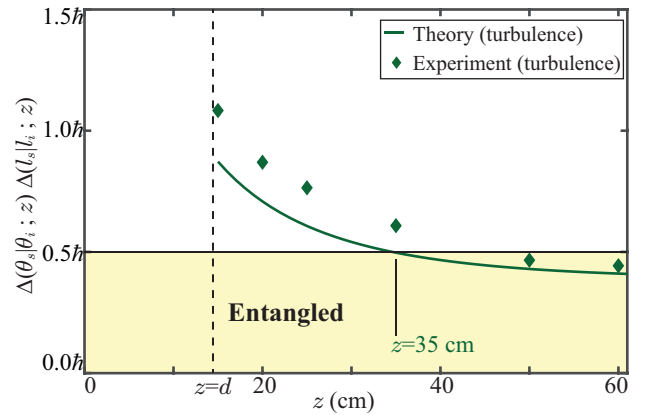}
\caption{Conditional angle-OAM uncertainty product $\Delta(\theta_s|\theta_i; z)\Delta(l_s|l_i; z)$ as a function of the propagation distance $z$ in the presence of a turbulent medium. As indicated on the plot, in the presence of turbulence, the theoretical prediction for entanglement revival is at $z=35$ cm while we observe it experimentally at about $45$ cm.}\label{FIG4}
\end{figure}

Next, we report our measurements of the condition uncertainty products $\Delta(y_s|y_i; z)\Delta(p_{ys}|p_{yi}; z)$ and $\Delta(\theta_{s}|\theta_{i}; z)\Delta(l_s|l_i; z)$ at various $z$ values in the absence of turbulence. We note that the conditional momentum and OAM uncertainties $\Delta(p_{ys}|p_{yi}; z)$ and $\Delta(l_s|l_i; z)$ remain constant as a function of $z$ due to the conservations of momentum and OAM, respectively, in SPDC. As a result, the functional dependence of  $\Delta(y_s|y_i; z)\Delta(p_{ys}|y_{yi}; z)$ and $\Delta(\theta_s|\theta_i; z)\Delta(l_{s}|l_{i}; z)$ on $z$ is same as that of $\Delta(y_s|y_i; z)$ and $\Delta(\theta_s|\theta_i; z)$, respectively. In our experiments, we measure $\Delta(p_{ys}|p_{yi}; z)$ to be $2.13\hbar$ mm$^{-1}$, which is in good agreement with the value $1.97\hbar$ mm$^{-1}$ calculated using Eq.~(\ref{eqn_spatial_PD}). We note that with our theoretical modelling based on the Gaussian pump beam, $\Delta(l_s|l_i; z)$ should be zero. However, because of the imperfection in the spatial profile of the beam and other background issues, $\Delta(l_s|l_i; z)$ is always finite in realistic experimental situations. We measure $\Delta(l_s|l_i; z)$ to be $0.72\hbar$ radian$^{-1}$ and use this in our experiments (See Sec.~II of the Supplementary Information for more details). Finally, we measure $\Delta(y_s|y_i; z)$ and $\Delta(\theta_{s}|\theta_{i}; z)$ at various $z$ values and plot the conditional uncertainty products $\Delta(y_s|y_i; z)\Delta(p_{ys}|p_{yi}; z)$ and $\Delta(\theta_s|\theta_i; z)\Delta(l_{s}|l_{i}; z)$ as a function of $z$ in  Figs.~\ref{FIG3}(a) and ~\ref{FIG3}(b), respectively.  Figures~\ref{FIG3}(a) and  ~\ref{FIG3}(b) also show the theory plots calculated using Eqs.~(\ref{eqn_spatial_PD}) and (\ref{eqn_angular_PD}). We see good agreement bewteen the theory and experiments. We note that it is relatively easier to measure $\Delta(y_s|y_i; z)\Delta(p_{ys}|y_{yi}; z)$ and $\Delta(\theta_s|\theta_i; z)\Delta(l_{s}|l_{i}; z)$ close to the crystal ($z<3$ mm) or in the far field $(z>10$ mm). However, due to the signal-to-noise limitations of the EMCCD camera, it is not possible to make measurements in the intermediate regions. (For details on these measurements, see Sec.~IV and V of the Supplementary Information.) We find that entanglements in both position-momentum and angle-OAM bases are lost within a few centimeters from the down-conversion crystal. However, while the position-momentum entanglement never revives, the angle-OAM entanglement, as calculated theoretically, revives after the photons have propagated 22 cm away from the crystal; experimentally, we find this distance to be about 28 cm. After the revival, the angle-OAM entanglement does not decay back.

We next investigate if the propagation-induced entanglement revival takes place in turbulence environments, which is quite often the limiting factor in the practical implementations of many entanglement-based applications. For this, we repeat our experiments depicted in Figs.~\ref{fig_setup1}(a) and \ref{fig_setup1}(b) in the angle-OAM basis with the blower heater (BH) switched on and kept at $z=15$ cm to introduce turbulence in the path of the down-converted photons. 
We experimentally measure the product $\Delta(\theta_s|\theta_i; z)\Delta(l_s|l_i; z)$ at different propagation distances ranging from $z=15$ cm to $z=60$ cm and plot them in Fig.~\ref{FIG4}. The solid line represents the theoretically calculated value of the uncertainty product. (For the detailed theoretical modelling of turbulence and the calculations of $\Delta(\theta_s|\theta_i; z)\Delta(l_s|l_i; z)$ in the presence of turbulence, see Sec.~VI and VII of the Supplementary Information.) Our theory results show that in the presence of turbulence, the angle-OAM entanglement revives at  $z=35$ cm; experimentally, we find this distance to be about 45 cm. Therefore, we find that although turbulence does affect angle-OAM entanglement in an adverse manner, its effect can be completely bypassed by just propagating the photons further ahead by some distance.

In conclusion, using the two-photon field produced by SPDC, we have reported experimental observations of propagation-induced entanglement revival in the angle-OAM bases. We have demonstrated entanglement revival even in the presence of turbulence, the only effect of which is to increase the propagation distance for revival. Once revived, the two photons remain entangled upto an arbitrary propagation distance. We note that the entanglement revival strategies in turbulence or random media are usually based on adaptive optics techniques \cite{valencia2020natphot, defienne2018prl1, lib2020sciadv}, which are usually based on the feedback mechanism and as a result quite difficult to implement. On the other hand, in our work, we have shown that the entanglement can be revived simply by further propagating the two-photon field by some distance, without having to use any adaptive optics techniques. Thus, unlike the position-momentum bases, the angle-OAM bases brings in an independent parameter---the propagation distance---for entanglement revival in turbulent environments and can therefore have important implications for long-distance quantum information applications.

\section{Author Contributions}

A.B. proposed the idea. A.B., M.K.J and A.K.J. worked out the theory and performed the experiments. S.K. derived many of the theoretical results and also carried out several numerical simulations. J.L. helped with coincidence counting basics. A.B. and A.K.J wrote the manuscript with inputs from all the co-authors. A.K.J. supervised the entire work.

\section{Acknowledgements}
We thank Sanjana Wanare for help with taking data remotely during the pandemic time. We acknowledge financial support through the research grant no. DST/ICPS/QuST/Theme-1/2019 from the Department of Science and Technology, Government of India, and the support of the EPSRC grant EP/T00097X/1. S.K. acknowledges the UGC, Government of India for financial support.

\bibliography{ref}

\end{document}


\title{{Supplementary information}: Propagation-induced entanglement revival}

\author{Abhinandan Bhattacharjee$^{1}$, Mritunjay K. Joshi$^{1}$, Suman Karan$^{1}$, Jonathan Leach$^{2}$, and Anand K. Jha$^{}$}

\email{akjha9@gmail.com}

\affiliation{$^{}$ Department of Physics, Indian Institute of Technology Kanpur, Kanpur, UP 208016, India\\
$^{2}$School of Engineering and Physical Sciences, Heriot-Watt University, Edinburgh, EH14 4AS, UK}

\date{\today}

\maketitle

\section{Calculation of conditional position and angle uncertainties}

\subsection{Derivation of the formulas}

For a Gaussian pump with beam waist at the crystal plane $z=0$, the two-photon wavefunction in the position basis at the crystal plane $z=0$ is given by \cite{edgar2012natcom,walborn2010phyrep,schneeloch2016jop}:
\begin{align}
\psi(\bm{\rho}_s, \bm{\rho}_i;0) = A \exp\left[-\frac{|\bm{\rho}_s+\bm{\rho}_i|^2}{4 w_0^2}\right]\exp\left[-\frac{|\bm{\rho}_s-\bm{\rho}_i|^2}{4\sigma_0^2}\right], \label{wavefunction at crystal}
\end{align}
where $\bm{\rho}_s \equiv (x_s,y_s)$ and $\bm{\rho}_i \equiv (x_i,y_i)$ are the transverse positions of the signal and idler photons, respectively at $z$, $k=\pi/\lambda_p$, and $|\bm{\rho}_s|=\rho_s$, etc. Also, $w_0$ is the pump beam waist at $z=0$, $\sigma_0=\sqrt{{0.455 L \lambda_p}/{2\pi}}$, $L$ is the length of the crystal, and $\lambda_p$ is the wavelength of the pump field. Using the two-photon wave-function $\psi(\bm{\rho}_s, \bm{\rho}_i;0)$ at $z=0$, we calculate the two-photon wave-function at $\psi(\bm{\rho}_s, \bm{\rho}_i;z)$ at $z$ and thereby the two-photon position probability distribution function $P(\bm{\rho}_s,\bm{\rho}_i;z)=|\psi(\bm{\rho}_s, \bm{\rho}_i;z)^* \psi(\bm{\rho}_s, \bm{\rho}_i;z)|$ at  $z$, which can be shown to be 
\begin{align}
P(\bm{\rho}_s,\bm{\rho}_i;z)=|A|^2 \exp\left[-\frac{|\bm{\rho}_s+\bm{\rho}_i|^2}{2w(z)^2}\right] \exp\left[-\frac{|\bm{\rho}_s-\bm{\rho}_i|^2} {2\sigma(z)^2}\right], \label{eqn_spatial_PD}
\end{align}
where $w(z)=w_0\sqrt{1+z^2/(k^2 w_0^4)}$ and $\sigma(z)=\sigma_0\sqrt{1+z^2/(k^2\sigma_0^4)}$.

\begin{figure*}[t!]
\includegraphics[scale=0.95]{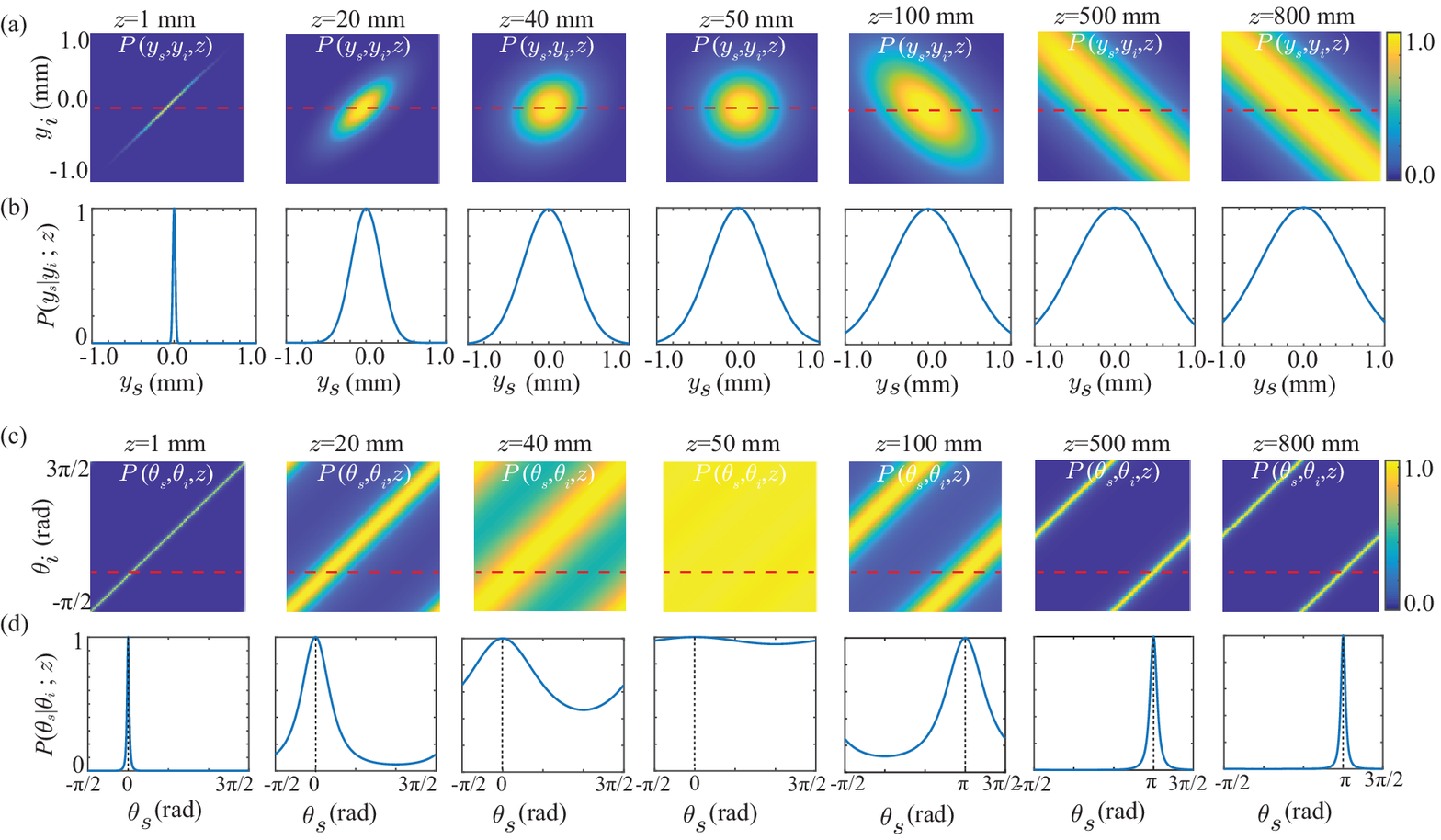}
\caption{(a) and (c) show the two-photon position probability distribution function $P(y_s,y_i;z)$ and the angle probability distribution function $P(\theta_s,\theta_i;z)$ respectively at various $z$ values. (b) and (d) show the  conditional position probability distribution function $P(y_s|y_i;z)$  and the conditional angle probability distribution function $P(\theta_s|\theta_i;z)$ of the signal photon at various $z$ values. 
}\label{fig1}
\end{figure*}

In addition to the position-momentum bases the down-converted photons are rendered entangled in the angle-OAM bases as well. We obtain the two-photon angle probability distribution by first writing $P(\bm{\rho}_s,\bm{\rho}_i;z)$ of Eq.~(\ref{eqn_spatial_PD}) in the polar coordinates using the transformations $\bm{\rho}_s=(r_s\cos\theta_s,r_s\sin\theta_s)$ and $\bm{\rho}_i=(r_i\cos\theta_i,r_i\sin\theta_i)$, where $(r_s,\theta_s)$ and $(r_i,\theta_i)$ are the polar coordinates of the signal and idler photons at $z$, etc. We therefore get:
\begin{align}
&|\bm{\rho}_s+\bm{\rho}_i|^2=r_s^2+r_i^2+2 r_s r_i \cos(\theta_s-\theta_i) \notag \\
&|\bm{\rho}_s-\bm{\rho}_i|^2=r_s^2+r_i^2-2 r_s r_i \cos(\theta_s-\theta_i) \notag \\
&P(r_s, \theta_s,r_i,\theta_i;z)=|A|^2 \exp\left[-\frac{r_s^2+r_i^2+2 r_s r_i \cos(\theta_s-\theta_i)}{2w(z)^2}\right] \notag \\
&\qquad\qquad\quad\times\exp\left[-\frac{r_s^2+r_i^2-2 r_s r_i \cos(\theta_s-\theta_i)}{2\sigma(z)^2}\right]\label{eqn_angular_PD-0}
\end{align}
We then integrate $P(r_s, \theta_s,r_i,\theta_i;z)$ over the radial coordinates in order to obtain the two-photon angle probability distribution function $P(\theta_s,\theta_i; z)$ as:
\begin{equation}
P(\theta_s,\theta_i;z)=\iint r_s r_i P(r_s,\theta_s,r_i,\theta_i;z) dr_s dr_i,
\label{eqn_angular_PD}
\end{equation}
Now, using the relevant experimental parameters $w_0=507$ $\mu$m, $L=5$ mm, and $\lambda_p=355$ nm in Eqs.~(\ref{eqn_spatial_PD}), and ~(\ref{eqn_angular_PD}), we calculate the two-photon position probability distribution $P(y_s,y_i;z)$ and the two-photon angle probability distribution $P(\theta_s,\theta_i;z)$ at different propagation distances $z$ as shown in Figs.~\ref{fig1}(a) and ~\ref{fig1}(c) respectively. In plottting $P(y_s,y_i;z)$ and $P(\theta_s,\theta_i;z)$ in Figs.~\ref{fig1}(a) and ~\ref{fig1}(c), we scale them in order to make their maximum values equal to one. We next calculate the corresponding conditional position probability distribution function $P(y_s|y_i;z)$ and the angle probability distribution function $P(\theta_s|\theta_i;z)$ by fixing $y_i=0$ mm in $P(y_s,y_i; z)$ and $\theta_i=0$ in $P(\theta_s,\theta_i; z)$.  Figures~\ref{fig1}(b) and \ref{fig1}(d) show $P(y_s|y_i; z)$ and $P(\theta_s|\theta_i; z)$ respectively. Figures~\ref{fig1}(a) and ~\ref{fig1}(b) show that in the near-field region the two down-converted photons have the maximum probability of arriving at the same transverse position. This is referred to as the position-correlation in the near-field region \cite{howell2004prl}. As the photon pair propagate away from the crystal plane, they become anti-correlated in position. Figures.~\ref{fig1}(c) and ~\ref{fig1}(d) show how the correlations in the angle basis change as a function of $z$. We find that in the near field, the signal and idler photons have the maximum probability of arriving at the same angular positions. However, in the far field, the two photons are most likely to arrive at antipodal locations, that is, at angles separated by $\pi$ radians. The standard deviations of $P(y_s|y_i; z)$, and $P(\theta_s|\theta_i; z)$ are referred to as the conditional position uncertainty $\Delta(y_s|y_i; z)$ and the conditional angle uncertainty $\Delta(\theta_s|\theta_i; z)$, respectively. From the plots in Fig.~\ref{fig1}, we find that as the down-converted photons propagate away from the crystal, the conditional position uncertainty increases monotonically. However, the conditional angle uncertainty increases initially but later begins to decrease monotonically. We use the conditional uncertainties $\Delta(y_s|y_i; z)$ and $\Delta(\theta_s|\theta_i; z)$ calculated in this section to compare with the experimentally measured values reported in Sec.~\ref{measure}. 

As calculated using Eqs.~(\ref{eqn_spatial_PD}) and (\ref{eqn_angular_PD}), the  conditional position and angle uncertainties behave differently upon propagation. Although it is very difficult to derive the general analytical expressions for the  conditional position and angle uncertainties, we have obtained analytical expressions for how the conditional position and angle uncertainties scale with $z$ in the near- and far-field regions.

\subsection{Near- and far-field behaviours of the conditional position uncertainty}

\begin{figure*}[t!]
\includegraphics[scale=1.0]{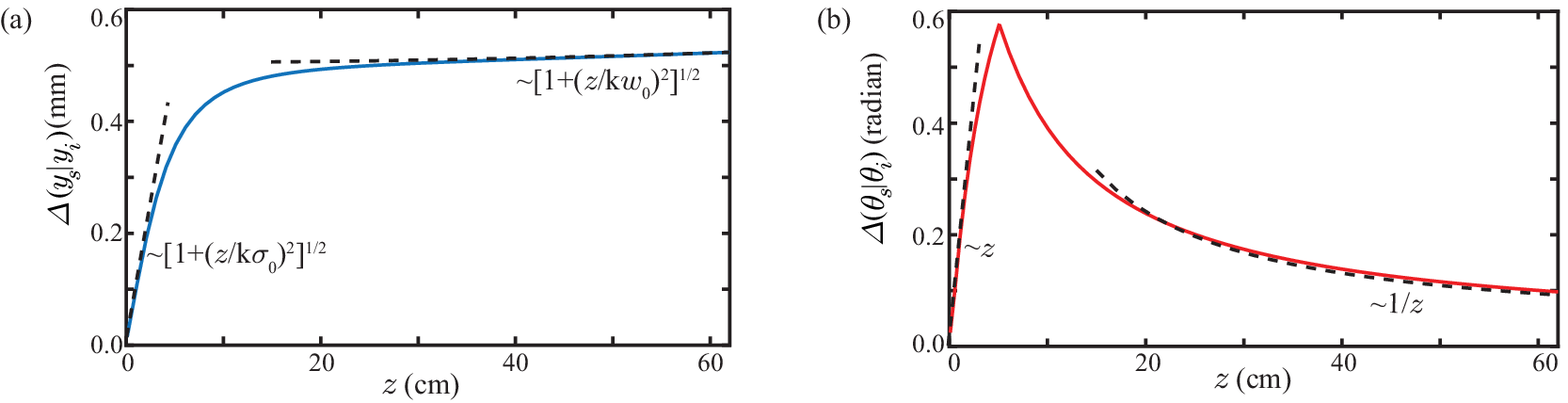}
\caption{(a) Numerically calculated conditional position uncertainty $\Delta(y_s|y_i; z)$ as a function of $z$. The two dotted lines show the $z$-scaling of the uncertainty in the near- and far-field regions. (b) Numerically calculated conditional angle uncertainty $\Delta(\theta_s|\theta_i; z)$ as a function of $z$. The two dotted lines show the $z$-scaling of the uncertainty in the near- and far-field regions. 
}\label{scaling}
\end{figure*}

The two-photon position probability distribution function is given by Eq.~(\ref{eqn_spatial_PD}). By setting ${\bm\rho_i=0}$, we write the conditional positional probability distribution function $P({\bm\rho_s}|{\bm\rho_i};z)$ as 
\begin{align}
&P(\bm{\rho}_s|\bm{\rho}_i;z)=|A|^2 \exp\left[-\frac{\rho_s^2}{2}\left(\frac{1}{2w(z)^2}+\frac{1}{2\sigma(z)^2}\right)\right], \notag \\
&{\rm where} \qquad  w(z)^2=w_0^2\left[1+\frac{z^2}{k^2 w_0^4}\right], \notag \\
& \qquad \qquad
\sigma(z)^2=\sigma_0^2 \left[1+\frac{z^2}{k^2 \sigma_0^4}\right],
 \label{eqn_spatial_PD2}
\end{align}
and $|\bm{\rho}_s|^2=\rho_s^2$. From Eq.~(\ref{eqn_spatial_PD2}), we obtain the conditional position uncertainty in the $y$-direction as
\begin{align}
\Delta(y_s|y_i; z)=\sqrt{\frac{1}{\frac{1}{2w(z)^2}+\frac{1}{2\sigma(z)^2}}}.
\end{align}
For the experimental parameters of interest, we have $w_0=507$ $\mu$m and $\sigma_0=11.3$ $\mu$m. Therefore, in the near-field region, we have $w(z)\gg \sigma(z)$ and thus  the conditional uncertainty in the $y$-direction becomes 
\begin{align}
\Delta(y_s|y_i; z)\approx \sigma(z)=\sigma_0 \sqrt{1+\frac{z^2}{k^2 \sigma_0^4}}. \label{position-near}
\end{align}
In the far-field, we have  $w(z)\ll \sigma(z)$ and thus the conditional uncertainty in the $y$-direction becomes 
\begin{align}
\Delta(y_s|y_i; z)\approx w(z)=w_0 \sqrt{1+\frac{z^2}{k^2 w_0^4}}. \label{position-far}
\end{align} 
From Eqs.~(\ref{position-near}) and (\ref{position-far}),  we find that the conditional position uncertainty $\Delta(y_s|y_i; z)$ increases monotonically as a function of $z$ in both the near- and far-field regions. While the uncertainty increases as $\sigma(z)$ in the near-field, it  increases as $w(z)$ in the far-field. Figure \ref{scaling}(a) shows the plot of the numerically calculated conditional position uncertainty $\Delta(y_s|y_i; z)$ as a function of $z$. The two dotted lines in Fig.~\ref{scaling}(a) show the $z$-scaling of the uncertainty in the near- and far-field regions. 

\subsection{Near- and far-field behaviours of the conditional angle uncertainty}

The two-photon angle probability distribution function is given by Eq.~(\ref{eqn_angular_PD}). The conditional angle probability distribution function $P(\theta_s|\theta_i; z)$ is obtained by setting $\theta_i=0$ in Eq.~(\ref{eqn_angular_PD}). As we are interested only in obtaining the near- and far-field scaling of the conditional angle uncertainty, we take $P(r_s,\theta_s,r_i,\theta_i;z)=P(r_s,\theta_s,\theta_i;z)\delta(r_s-r_i)$. Thus we write Eq.~(\ref{eqn_angular_PD}) as
\begin{equation}
P(\theta_s,\theta_i;z)=\int P(r,\theta_s,\theta_i;z) r^2 dr.
\end{equation}
Using Eq.~(\ref{eqn_angular_PD-0}) and the Mathematica software, we evaluate the above integral and obtain 
\begin{equation}
P(\theta_s,\theta_i;z)=\frac{P_0}{[C+D \cos(\theta_s-\theta_i)]^{3/2}},
\end{equation}
where
\begin{align}
& P_0=\frac{|A|^2\sqrt{\pi/2}}{8}, \notag \\
& C=\frac{1}{2}\left[\frac{1}{w(z)^2}+\frac{1}{\sigma(z)^2}\right], \notag \\
& D=\frac{1}{2}\left[\frac{1}{w(z)^2}-\frac{1}{\sigma(z)^2}\right]. 
\end{align}
The ratio of $C$ and $D$ can be written as
\begin{align}
\frac{C}{D}&=\frac{w(z)^2+\sigma(z)^2}{-w(z)^2+\sigma(z)^2} \notag \\
&=\frac{(w_0^2+\sigma_0^2)+\frac{z^2}{k^2} \left[\frac{1}{w_0^2}+\frac{1}{\sigma_0^2}\right] }{(-w_0^2+\sigma_0^2)+\frac{z^2}{k^2} \left[-\frac{1}{w_0^2}+\frac{1}{\sigma_0^2}\right]}. \label{ratio}
\end{align}
In our experiments, we have $w_0=507$ $\mu$m and $\sigma_0=11$ $\mu$m. Thus we have $w_0\gg\sigma_0$, and under this approximation we write the above ratio as
\begin{align}
\frac{C}{D}=\frac{w_0^2+z^2/(k^2 \sigma_0^2) }
{-w_0^2+z^2/(k^2 \sigma_0^2) } = \frac{z^2+k^2\sigma_0^2w_0^2}{z^2-k^2\sigma_0^2w_0^2}.
\end{align}
Next, we study the behaviour of $P(\theta_s,\theta_i;z)$ in the near field regions. We make use of the fact that for $\theta_i=0$, $P(\theta_s,\theta_i;z)$ is maximum at $\theta_s=0$. Therefore, we have
\begin{align}
P^{\rm max}(\theta_s,\theta_i=0;z)=\frac{P_0}{[C+D]^{3/2}}.
\end{align}
We next find the value of $\theta_s$ at which $P(\theta_s,\theta_i;z)=P^{\rm max}(\theta_s,\theta_i=0;z)/2$, in which case $\theta_s$ can be taken as the half-width of the conditional angle probability distribution function. We thus equate
\begin{align}
& P(\theta_s,\theta_i;z)=P^{\rm max}(\theta_s,\theta_i=0;z)/2 \notag \\
{\rm or,} \quad & \frac{P_0}{[C+D\cos\theta_s]^{3/2}} =\frac{P_0}{2[C+D]^{3/2}} \notag \\
{\rm or,} \quad & C+D\cos\theta_s = 2^{2/3}(C+D)
\end{align}
Solving the above equation, we get two solutions for $\theta_s$:
\begin{align}
&\theta_s^{(+)}=\cos^{-1}\left[(2^{2/3}-1)\frac{C}{D}+2^{2/3}\right] \notag \\
{\rm and} \qquad&\theta_s^{(-)}=-\cos^{-1}\left[(2^{2/3}-1)\frac{C}{D}+2^{2/3}\right].
\end{align}
The angle uncertainty $\Delta(\theta_s|\theta_i; z)$ can therefore be written as
\begin{align}
\Delta(\theta_s|\theta_i; z)&=\theta_s^{(+)}-\theta_s^{(-)} \notag \\ &=2\cos^{-1}\left[(2^{2/3}-1)\frac{C}{D}+2^{2/3}\right].
\end{align}
Using the approximation $\cos^{-1}x=\sqrt{2(1-x)}$ for $x\in[0,1]$, we write the above uncertainty as:
\begin{align}
\Delta(\theta_s|\theta_i; z)=2\sqrt{2(2^{2/3}-1)\left[-\frac{C}{D}-1\right]}.
\end{align}
Substituting for $C/D$ from Eq.~(\ref{ratio}), we obtain
\begin{align}
\Delta(\theta_s|\theta_i; z)=4\sqrt{2^{2/3}-1}\times \sqrt{\frac{z^2}{k^2\sigma_0^2w_0^2 -z^2}}.
\end{align}
In the near-field regions, we have $k^2\sigma_0^2w_0^2 \gg z^2$. Therefore, we can write the angle uncertainty in the near-field regions as 
\begin{align}
\Delta(\theta_s|\theta_i; z)\approx\frac{4\sqrt{2^{2/3}-1}}{k\sigma_0w_0} z. \label{angle-near}
\end{align}
Thus in the near-field regions the angle uncertainty increases linearly with $z$. In the far-field, we use the fact that for $\theta_i=0$, $P(\theta_s,\theta_i;z)$ is maximum at $\theta_s=\pi$. Therefore, in the far-filed we have $ P^{\rm max}(\theta_s,\theta_i=0;z)=P_0/[C-D]^{3/2}$. Now, proceeding in the similar manner as above, and using the far-field approximation $z^2 \gg k^2\sigma_0^2w_0^2$, we find the angle uncertainty in the far-field regions to be
\begin{align}
\Delta(\theta_s|\theta_i; z)\approx 4\sqrt{2^{2/3}-1} k\sigma_0w_0 \frac{1}{z}   \label{angle-far}
\end{align}
We thus find that in the far-field region the angle uncertainty $\Delta(\theta_s|\theta_i; z)$ becomes inversely proportional to $z$ and as a consequence decreases upon propagation. Figure \ref{scaling}(b) shows the numerically calculated conditional angle uncertainty $\Delta(\theta_s|\theta_i; z)$ as a function of $z$. The two dotted lines in Fig.~\ref{scaling}(b) show the $z$-scaling of the uncertainty in the near- and far-field regions.

\section{Calculation of conditional momentum and OAM uncertainties}\label{app-sec1}

Using the two-photon wave-function in the position basis $\psi(\bm{\rho}_s, \bm{\rho}_i;z)$ calculated in the previous section, we calculate the two-photon wave-function $\psi(\bm{\rho}_s, \bm{\rho}_i;z)$ in the transverse momentum basis, which is given by  \cite{schneeloch2016jop}
\begin{multline}
	\psi(\bm{p}_s, \bm{p}_i;z) = A \exp\left[-\frac{|\bm{p}_s+\bm{p}_i|^2w_0^2}{4\hbar^2}\right] \\ \times \exp\left[-\frac{|\bm{p}_s-\bm{p}_i|^2\sigma_0^2}{4\hbar^2}\right]\exp\left[-\frac{iz}{k\hbar^2}|\bm{p}_s^2+\bm{p}_i^2|\right],
	\label{eqn_two_photon_mom_wf}
\end{multline} 
where $\bm{p}_s \equiv (p_{sx},p_{sy})$ and $\bm{p}_i \equiv (p_{ix},p_{iy})$ are the transverse momenta of the signal and idler photons, respectively. Using the above equation, we find that the conditional momentum probability distribution function $P(\bm{p}_s|\bm{p}_i; z)$ of the signal photon is given by
\begin{equation}
P(\bm{p}_s|\bm{p}_i;z) = A \exp\left[-\frac{p_s^2(w_0^2+\sigma_0^2)}{2\hbar^2}\right],
\label{eqn_two_photon_mom_prob}
\end{equation} 
where $p_s^2=|\bm{p}_s|^2$. The standard deviation of $P(\bm{p}_s|\bm{p}_i; z)$ in the $y$-direction $\Delta(p_{sy}|p_{iy}; z)$ is the conditional momentum uncertainty of the signal photon. The above equation shows that $P(\bm{p}_s|\bm{p}_i; z)$ is independent of $z$ and that $\Delta(p_{sy}|p_{iy}; z)$ does not change upon propagation.

For a Gaussian pump the two-photon state produced by SPDC in the OAM basis can be written as \cite{kulkarni2017natcom,leach2010science}
\begin{equation}
	|\Psi\rangle = \sum_{l_s=-\infty}^{\infty} \sqrt{S_{l_s}} |l_s\rangle_s |-l_s\rangle_i,
	\label{eqn_two_photon_oam}
\end{equation}
where $l_s\hbar$ and $-l_s\hbar$ are the OAMs of signal and idler photons, respectively. The form of the two-photon state above implies that if the signal photon is detected with OAM $l_s\hbar$, then the idler photon is guaranteed to be detected with OAM $-l_s\hbar$. For the above state, and with $l_i=0$, the conditional two-photon OAM probability distribution function takes the following form: $P(l_s|l_i; z)=S_{l_s} \delta_{l_s, 0}$. This implies that the corresponding conditional OAM uncertainty $\Delta(l_s|l_i; z)$ is equal to zero. However, in an experimental situation, one always measures $\Delta(l_s|l_i; z)$ to be non-zero \cite{leach2010science}. There are several reasons for this, which includes the pump not being an ideal Gaussian beam, the experimental imperfections such as misalignment and background noise, and the mode dependent detection efficiencies of OAM detectors. These cause an additional contribution in $P(l_s|l_i; z)$ measurement. Therefore, in our experiments, we model the conditional OAM probability distribution function as:
\begin{equation}
P(l_s|l_i; z)=S_{l_s} \delta_{l_s, 0} + N \exp\left[-\frac{l_s^2}{2 \sigma_f^2}\right], 
\label{eqn_two_photon_oam_1}
\end{equation}
where $S_0$, $\sigma_f$ and $N$ are the fitting parameters. We take the width of $P(l_s|l_i; z)$ as the conditional OAM uncertainty $\Delta(l_s|l_i; z)$.

\section{Coincidence Measurement with EMCCD camera}

In this section we outline how we use an Andor iXon Ultra-897 EMCCD camera having $512 \times 512$ pixel grid with $16 \times 16$ $\mu$m$^2$ pixel-size for measuring  coincidence counts in position and angle bases. For this, we record $10^6$-$10^7$ images of the SPDC field with an exposure time of $1$ ms - $5$ ms over a few hours with average flux of $0.5$ - $2.0$ photons per pixel. We operate the camera at -$60^{\degree}$C with the electron-multiplication gain of $1000$, the horizontal pixel readout rate of $5$-$17$ MHz, the vertical pixel shift speed of $0.3$ $\mu$s, and the vertical clock amplitude of $+4$V. In SPDC, a signal and idler photon pair gets generated within a very short time interval, usually of the order of $100$ fs, which is much smaller than the exposure time ($1-5$ ms) of the EMCCD camera. Therefore, in all likelihood, the signal and idler photons belonging to a pair arrive within the same image. However, within the same image, we can also have signal and idler photons that are not from the same down-conversion pair. These give rise to the accidental coincidences, which between the pixels (or pixel groups) $p$ and $q$ can be estimated by computing the coincidence counts  between $k^{th}$ and $(k+1)^{th}$ images. Therefore, as detailed in Ref.~\cite{reichert2018scirpt,defienne2018prl}, the true coincidence count $C_{pq}$ between two pixels or two groups of pixels, $p$ and $q$, of the EMCCD camera can be expressed as
\begin{align}
	C_{pq} &= \frac{1}{N}\sum_{k=1}^{N}n_p^{(k)}n_q^{(k)} - \frac{1}{N}\sum_{k=1}^{N}n_p^{(k)}n_q^{(k+1)},
	\label{eqn_final_coincidence coint}
\end{align}
where the first term is the coincidence due to down-converted pairs and the second term is the accidental coincidence.

\begin{figure}[t!]
\centering
\includegraphics[scale=1]{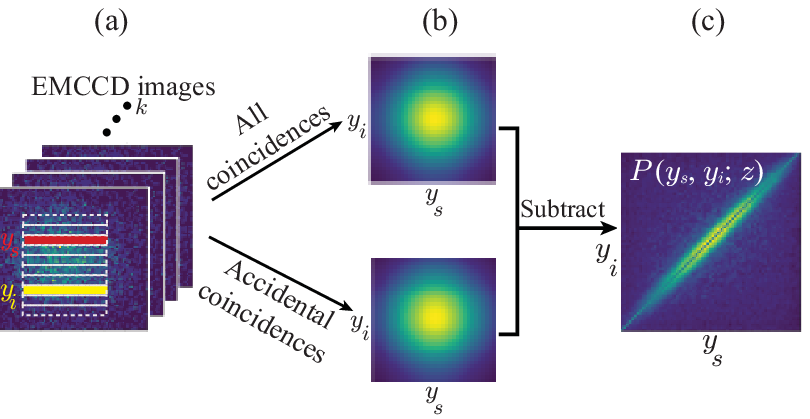}
\caption{(a) Acquired images of SPDC field and binning the pixels into signal  $y_s$ and idler $y_i$ bars. (b) The top and the bottom images represent the total coincidence and the accidental coincidence calculated using the first and the second terms of Eq.~(\ref{eqn_final_coincidence coint_pos}), respectively. Subtraction of these two terms gives (c) the measured two-photon position probability distribution function $P(y_s,y_i; z)$.}
\label{fig_ROIpos}
\end{figure}

For measuring the two-photon position probability distribution function $P(y_s,y_i; z)$, we take millions of images using the EMCCD camera. For each image, we group the pixels into horizontal strips, $y_s$ and $y_i$, as shown in Fig.~\ref{fig_ROIpos}(a). The coincidence count between $y_s$ and $y_i$ can be written using Eq.~(\ref{eqn_final_coincidence coint}) as
\begin{figure}[t!]
\centering
\includegraphics[scale=1.0]{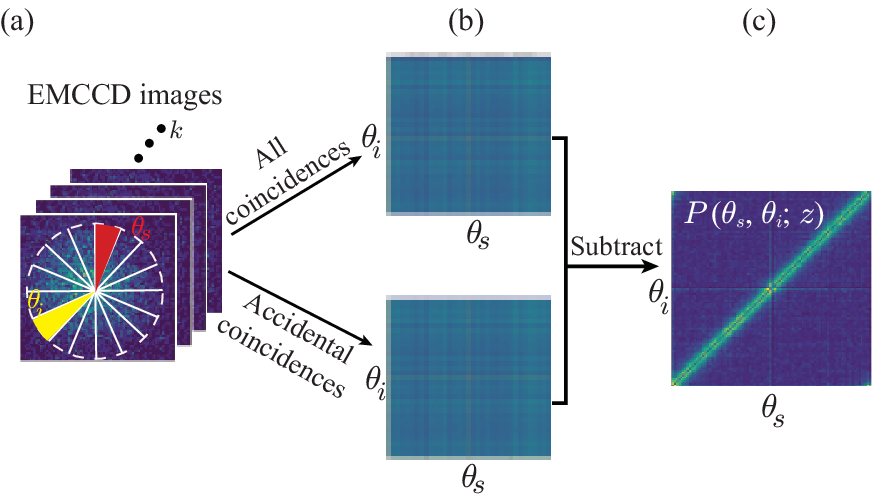}
\caption{(a) Acquired images of SPDC field and binning the pixels into signal angular sector $\theta_s$ and idler angular sector $\theta_i$. (b) The top and the bottom images represent the total coincidence and the accidental coincidence calculated using the first and the second terms of Eq.~(\ref{eqn_final_coincidence coint_ang}), respectively. Subtraction of these two terms gives (c) the measured two-photon angle probability distribution function $P(\theta_s,\theta_i; z)$.}
\label{fig_ROIang}
\end{figure}
\begin{equation}
C_{y_s y_i} = \frac{1}{N}\sum_{k=1}^{N}  n_{y_s}^{(k)} n_{y_i}^{(k)}  -  \frac{1}{N}\sum_{k=1}^{N} n_{y_s}^{(k)} n_{y_i}^{(k+1)},
\label{eqn_final_coincidence coint_pos}
\end{equation}
where $n_{y_s}^{(k)}$ and $n_{y_i}^{(k)}$ are the photon counts of $y_s$ and $y_i$  respectively. The top image in Fig.~\ref{fig_ROIpos}(b) represents total coincidence as a function of $y_s$ and $y_i$, and it is evaluated by using the first term of Eq.~(\ref{eqn_final_coincidence coint_pos}). The bottom image in Fig.~\ref{fig_ROIpos}(b) represents the accidental coincidence as a function of $y_s$ and $y_i$, and it is evaluated using the second term of Eq.~(\ref{eqn_final_coincidence coint_pos}). The difference of these two images is proportional to the true coincidence $C_{y_s y_i}$ and thus to the two-photon position probability distribution function $P(y_s,y_i; z)$, as shown in Fig.~\ref{fig_ROIpos}(c). At $y_s = y_i$, the correlation becomes artificially perfect because we are correlating a pixel with itself. So, we discard the points with $y_s = y_i$ as outliers.

\begin{figure*}[t!]
\includegraphics[scale=1.0]{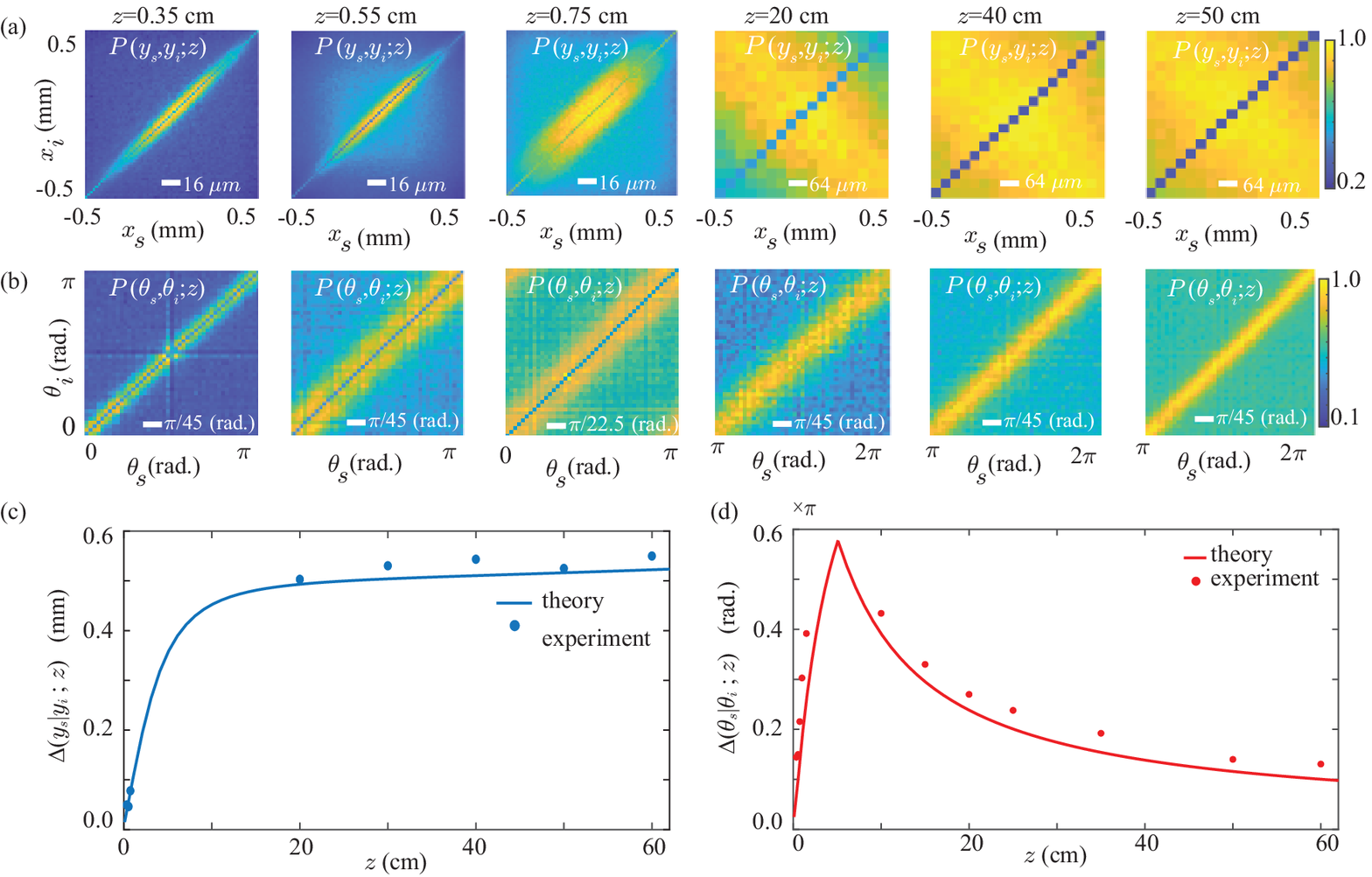}
\caption{(a) and (b) show the experimentally measured two-photon position and angle probability distribution $P(y_s,y_i; z)$ and $P(\theta_s,\theta_i; z)$ as a function of the propagation distance $z$. (c) and (d) show the plots of the position and angle uncertainties $\Delta(y_s|y_i; z)$ and $\Delta(\theta_s|\theta_i; z)$ as a function of $z$. The experimental points are shown with solid dots while the solid curve represents the theoretical predictions.}
\label{2d_plots}
\end{figure*}

For measuring the two-photon angle probability distribution $P(\theta_s,\theta_i; z)$, we group the pixels for each image into angular sectors as shown in Fig.~\ref{fig_ROIang}(a). The coincidence count between the angular sectors at $\theta_s$ and $\theta_i$ is given by:
\begin{equation}
	C_{\theta_s \theta_i} = \frac{1}{N}\sum_{k=1}^{N}  n_{\theta_s}^{(k)} n_{\theta_i}^{(k)}  -  \frac{1}{N}\sum_{k=1}^{N} n_{\theta_s}^{(k)} n_{\theta_i}^{(k+1)} ,
	\label{eqn_final_coincidence coint_ang}
\end{equation}
where $n_{\theta_s}^{(k)}$ and $n_{\theta_i}^{(k)}$ are the photon counts of angular sectors $\theta_s$ and $\theta_i$ respectively. The top image in Fig.~\ref{fig_ROIang}(b) represents the total coincidence as a function of $\theta_s$ and $\theta_i$, and it is evaluated using the first term of Eq.~(\ref{eqn_final_coincidence coint_ang}). The bottom image in Fig.~\ref{fig_ROIang}(b) represents the accidental coincidence as a function of $\theta_s$ and $\theta_i$, and it evaluated using the second term of Eq.~(\ref{eqn_final_coincidence coint_ang}). The difference of these two images is proportional to the two-photon angle probability distribution $P(\theta_s,\theta_i; z)$, as shown in Fig.~\ref{fig_ROIang}(c). At $\theta_s = \theta_i$, the correlation becomes artificially perfect because we are correlating a pixel with itself. So, we discard the points with $\theta_s = \theta_i$ as outliers.

\section{Measurement of the two-photon position and angle probability distribution function}\label{measure}

Figure~\ref{2d_plots}(a) shows the experimentally measured two-photon position probability distribution function $P(y_s,y_i;z)$ at different $z$. For imaging the transverse planes between $z=0.35$ cm and $z=1.5$ cm, we keep the magnification of the imaging system to be 1 while for imaging the transverse planes between $z=10$ and $z=60$ cm, we keep the magnification to be $0.25$. We scale the measured $P(y_s,y_i;z)$ such that its maximum value is equal to one. We find that the photons are correlated in position in the near field whereas they get position anti-correlated in the far-field. In order to extract $\Delta(y_s|y_i;z)$ from the measured $P(y_s,y_i;z)$, we fit $P(y_s,y_i;z)$ with the function: $P_f(y_s,y_i; z)=bP_r(y_s,y_i; z)+ aP_n(y_s,y_i; z)$, where $P_r(y_s,y_i; z)=\exp[-(y_s+y_i-d)^2/(2\sigma_1^2(z))] \times \exp[-(y_s-y_i-f)^2/(2\sigma_2^2(z))]$ is considered as the probability distribution due to the down-converted photons, while $P_n(y_s,y_i; z)=\exp[-(y_s+y_i-d)^2/(2n^2)] \times \exp[-(y_s-y_i-f)^2/(2 m^2)]$ is considered as the noise contribution. Here $b$, $a$, $\sigma_1(z)$, $\sigma_2(z)$, $d$, $f$, $m$ and $n$ are the fitting parameters. We consider $n\gg\sigma_1(z)$, $m\gg\sigma_2(z)$ such that the noise contribution remains much broader than the two-photon position probability distribution. The width $\Delta(y_s|y_i=0;z)$ can now be expressed as $\Delta(y_s|y_i=0;z)=\sigma_1(z)\sigma_2(z)/\sqrt{\sigma_1^2(z)+\sigma_2^2(z)}$. Figure~\ref{2d_plots}(c) shows $\Delta(y_s|y_i=0;z)$ as a function of $z$. The theory plot has been calculated using the expression given in Eq.~(\ref{eqn_spatial_PD}).

Figure~\ref{2d_plots}(b) shows the experimentally measured $P(\theta_s,\theta_i;z)$ at different $z$.  We scale $P(\theta_s,\theta_i;z)$ such that its maximum value is equal to one. The $P(\theta_s,\theta_i;z)$ plots show that near the crystal plane the signal and idler photons have the highest probability of arriving at the same angular positions. However in the far-field the two photons are most likely to arrive at angular positions separated by $\pi$ radians. We fit the measured $P(\theta_s,\theta_i;z)$ with the analytic function: $P_f(\theta_s,\theta_i; z)=bP_r(\theta_s,\theta_i; z)+a$, where $P_r(\theta_s,\theta_i;z)=1/(1+q\cos(\theta_s-\theta_i-c))^{3/2}$.  Here, $b$, $a$, $q$, and $c$, are the fitting parameters. We derive the fitting function by putting $r_s=r_i$ in Eq.~(\ref{eqn_angular_PD}). Next, we evaluate $\Delta(\theta_s|\theta_i;z)$ by finding the standard deviation of $P_r(\theta_s|\theta_i;z)$ at various $z$ values. Figure.~\ref{2d_plots}(d) shows the experimental $\Delta(\theta_s|\theta_i;z)$ as a function of $z$. We find that near the crystal $\Delta(\theta_s|\theta_i;z)$ increases as a function of $z$.  However, beyond $z=10$ cm, $\Delta(\theta_s|\theta_i;z)$ starts to monotonically decreas as a function of $z$. The theory plot has been calculated using the expression given in Eq.~(\ref{eqn_angular_PD}). We see a good match between the theory and experiments.

\section{Measurement of the two-photon momentum and OAM probability distribution function}

\begin{figure}[t!]
\centering
\includegraphics[scale=1.0]{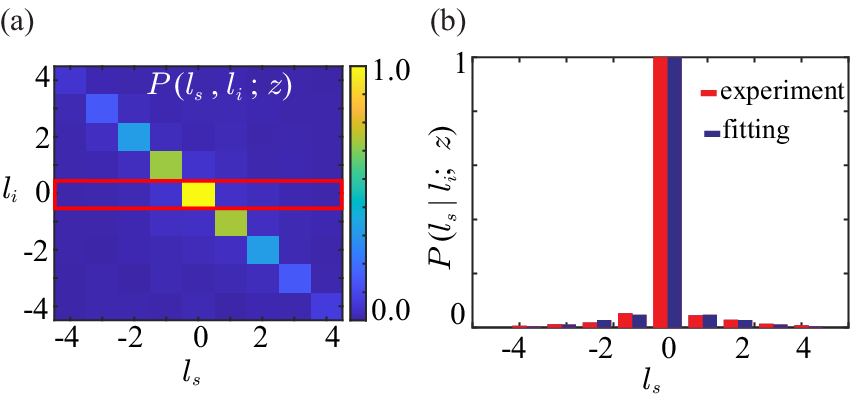}
\caption{(a) and (b) are the experimentally measured  $P(l_s, l_i; z)$ and $P(l_s|l_i; z)$ at $z=50$ cm. The fitting is  based on the noise model described in Eq.~(\ref{eqn_two_photon_oam_1}).}\label{OAM_plot}
\end{figure}

Equation (\ref{eqn_two_photon_mom_prob}) shows that $P(\bm{p}_s|\bm{p}_i; z)$ is independent of $z$ and that $\Delta(p_{sy}|p_{iy}; z)$ does not change upon propagation. For the given experimental parameters, the calculated value of $\Delta(p_{sy}|p_{iy}; z)$ is $1.97\hbar$ mm $^{-1}$. The two-photon OAM probability distribution function $P(l_s, l_i; z)$ remains unchanged as a function of $z$. We verify this by making several measurements of $P(l_s, l_i; z)$ as a function of $z$. We plot the experimentally measured two-photon OAM probability distribution function $P(l_s, l_i; z)$ and the conditional OAM probability distribution function $P(l_s|l_i; z)$ at $z=50$ cm in Figs.~\ref{OAM_plot}(a) and \ref{OAM_plot}(b), respectively. As described in Sec.~\ref{app-sec1}, we fit the conditional distribution with the analytical function $P(l_s|l_i; z)=S_{l_s} \delta_{l_s, 0}  + N \exp\left[-l_s^2 /(2 \sigma_f^2)\right]$,  where $S_0$, $N$ and $\sigma_f$ are the fitting parameters, and thus find the uncertainty  $\Delta(l_s|l_i;z)$ to be $0.72 \hbar$ in our experiments.

\section{Measurement of the two-photon angle probability distribution function in turbulence}\label{app-sec5}

\begin{figure*}[t!]
	\centering
    \includegraphics[scale=1.0]{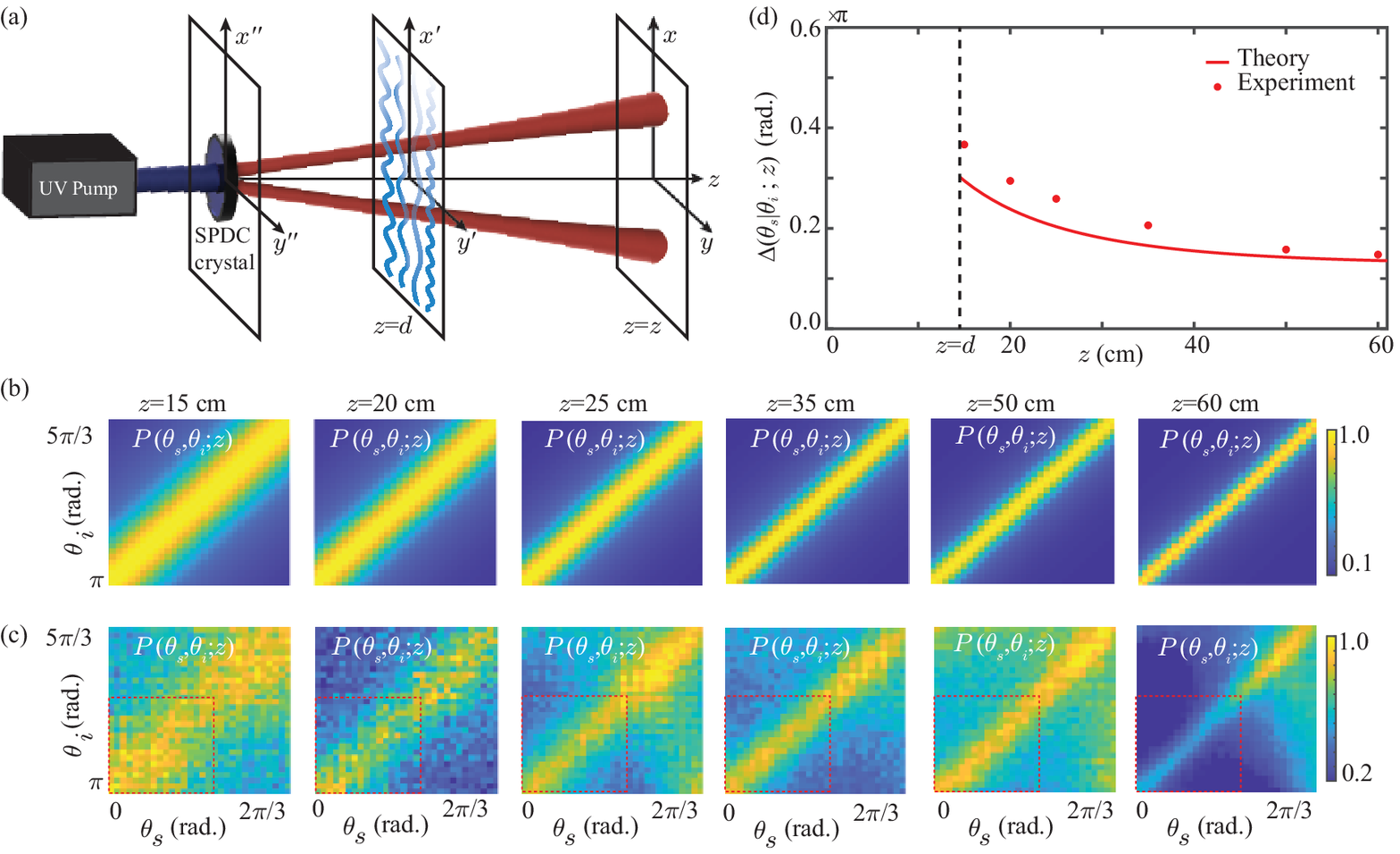}
    \caption{(a) Illustrating the propagation of the down-converted field in the presence of turbulence. (b) Theoretically calculated and (c) experimentally measured two-photon angle probability distribution function $P(\theta_s,\theta_i;z)$ at various $z$ in the presence of turbulence. (d) The theoretical and experimental plots of $\Delta(\theta_s|\theta_i;z)$ as a function of $z$.}
    \label{turb_plot}
\end{figure*}  

In this section, we present how the conditional angle uncertainty propagates after the two-photon field passes through turbulence. Figure~\ref{turb_plot}(a) illustrates the propagation of SPDC photons through a  planar turbulence kept at a distance $z=d$ from the crystal plane located at $z=0$. We are interested in finding the two-photon angle probability distribution function at a propagation distance $z$. The presence of turbulence introduces statistical randomness in the two-photon field, and so we need to describe the field propagation in terms of the two-photon cross-spectral density function. From $z=0$ up to $z=d$, the two photon field remains pure and can be described by the two-photon wave-function $\psi(\bm{\rho}_{s}, \bm{\rho}_{i};z)$. Therefore, the two-photon cross-spectral density function $W(\bm{\rho}'_{s1}, \bm{\rho}'_{i1},\bm{\rho}'_{s2}, \bm{\rho}'_{i2};z_t)$ right after the turbulence plane $z=d$ can be written as
\begin{align}
W(\bm{\rho}'_{s1}, \bm{\rho}'_{i1},&\bm{\rho}'_{s2}, \bm{\rho}'_{i2}; d) \notag\\
&= \psi^*(\bm{\rho}'_{s2}, \bm{\rho}'_{i2};d)\psi(\bm{\rho}'_{s1}, \bm{\rho}'_{i1};d) \notag\\
&\qquad\times W_{\rm turb}(\bm{\rho}'_{s1}, \bm{\rho}'_{s2}, \bm{\rho}'_{i1}, \bm{\rho}'_{i2}), 
\label{eqn_two_photon_turb}
\end{align}
where, $\bm{\rho}'_{s} \equiv (x'_{s}, y'_{s})$, and $\bm{\rho}'_{i} \equiv (x'_{i}, y'_{i})$ are the transverse co-ordinates of signal and idler photons respectively, at $z=d$. The term $\psi^*(\bm{\rho}'_{s2}, \bm{\rho}'_{i2};d)\psi(\bm{\rho}'_{s1}, \bm{\rho}'_{i1};d)$ is the two-photon cross spectral density function right before the turbulence plane. The effect due to the turbulence is captured through the cross-spectral density function, which we approximate by modelling the turbulence in terms of a Gaussian function: $W_{\rm turb}(\bm{\rho}'_{s1}, \bm{\rho}'_{s2}), \bm{\rho}'_{i1}), \bm{\rho}'_{i2})=\exp\left[-[|\bm{\rho}'_{s2}-\bm{\rho}'_{s1}|^2+|\bm{\rho}'_{i2}-\bm{\rho}'_{i1}|^2]/(2r^2)]\right]$, where $r$ is the turbulence strength \cite{avetisyan2016oe, boyd2011optexp}. We rewrite the above equation as
\begin{multline}
	W(\bm{\rho}'_{s1}, \bm{\rho}'_{i1},\bm{\rho}'_{s2}, \bm{\rho}'_{i2};d) = W_{\rm turb}(\bm{\rho}'_{s1}, \bm{\rho}'_{s2}, \bm{\rho}'_{i1}, \bm{\rho}'_{i2}) \\
	\times e^{\frac{ik}{2d}(\rho'^2_{s2}+\rho'^2_{i2}-\rho'^2_{s1}-\rho'^2_{i1})} 
	\int \psi^*(\bm{\rho}''_{s1}, \bm{\rho}''_{i1}; 0) \psi(\bm{\rho}''_{s2},\bm{\rho}''_{i2}; 0) \\
	\times e^{\frac{ik}{2d}(\rho''^2_{s2}+\rho''^2_{i2}-\rho''^2_{s1}-\rho''^2_{i1})} 
	e^{-\frac{ik}{d}(\bm{\rho}'_{s2}\cdot\bm{\rho}''_{s2}-\bm{\rho}'_{s1}\cdot\bm{\rho}''_{s1})} \\
	\times e^{-\frac{ik}{d}(\bm{\rho}'_{i2} \cdot \bm{\rho}''_{i2}-\bm{\rho}'_{i1} \cdot \bm{\rho}''_{i1})} d\bm{\rho}''_{s2}d\bm{\rho}''_{s1} d\bm{\rho}''_{i2}d\bm{\rho}''_{i1},
	\label{eqn_two_photon_turb_prob1}
\end{multline}
where $\psi(\bm{\rho}''_s, \bm{\rho}''_i;0)$ is the two-photon wave-function at the crystal plane $z=0$ and is given by Eq.~(\ref{wavefunction at crystal}). Now, by propagating  $W(\bm{\rho}'_{s1}, \bm{\rho}'_{i1},\bm{\rho}'_{s2}, \bm{\rho}'_{i2};d)$ from $z=d$ up to $z=z$, we find the two-photon cross-spectral density function at $z$ and thereby the two-photon position probability distribution function $P(\bm{\rho}_{s}, \bm{\rho}_{i};z)$:
\begin{multline}
P(\bm{\rho}_{s}, \bm{\rho}_{i};z) = \int W(\bm{\rho}'_{s1}, \bm{\rho}'_{i1},\bm{\rho}'_{s2}, \bm{\rho}'_{i2};z_t) e^{\frac{ik}{2(z-d)}(\rho'^2_{s2}+\rho'^2_{i2})} \\
\times e^{-\frac{ik}{2(z-d)}(\rho'^2_{s1}+\rho'^2_{i1})}e^{-\frac{ik}{(z-d)}\bm{\rho}_{s} \cdot (\bm{\rho}'_{s2}-\bm{\rho}'_{s1})}e^{\frac{ik}{(z-d)}\bm{\rho}_{i} \cdot (\bm{\rho}'_{i2}-\bm{\rho}'_{i1})} \\
\times d\bm{\rho}'_{s2}d\bm{\rho}'_{s1}d\bm{\rho}'_{i2}d\bm{\rho}'_{i1}.
\label{eqn_two_photon_turb_prob}
\end{multline}
By substituting Eq.~(\ref{eqn_two_photon_turb_prob1}) into Eq.~(\ref{eqn_two_photon_turb_prob}), we compute $P(\bm{\rho}_s,\bm{\rho}_i;z)$ as a function of $z$. We then use the transformations $\bm{\rho}_{s}=\left(r_{s} \cos \theta_{s}, r_{s} \sin \theta_{s}\right)$ and $\bm{\rho}_{i}=\left(r_{i} \cos \theta_{i}, r_{i} \sin \theta_{i}\right)$ in order to obtain $P\left(\theta_{s}, \theta_{i} ; z\right)$ using Eq.~(\ref{eqn_angular_PD}). 

Figure~\ref{turb_plot}(c) shows the experimentally measured $P(\theta_s,\theta_i;z)$ at different $z$ (see Sec.~\ref{measure} for the measurement details). Figure~\ref{turb_plot}(b) shows the $P(\theta_s,\theta_i;z)$ calculated using Eq.~(\ref{eqn_angular_PD}) at different $z$ for the relevant experimental parameters of $d=15$ cm, $L=5$ mm, $w_0=507$ $\mu$m. For the theoretical plots, we use the turbulence strength $r$ as a fitting parameter and find its value to be $0.125$ mm. We note that the experimentally measured $P(\theta_s,\theta_i;z)$ contains some noise distribution, which gets prominent at large $z$. This can be attributed to the fact that in the experiment, we insert a distributed turbulence in the path of the two-photon field, whereas in the theory, we approximate that as a planar turbulence. Nevertheless, the diagonal correlation in the experimentally observed $P(\theta_s,\theta_i;z)$ matches with the theoretical predictions. Using the procedure described in Sec.~\ref{measure} B, we extract the conditional angle uncertainty $\Delta(\theta_s|\theta_i;z)$ from Figure~\ref{turb_plot}(c) and plot them in Figure~\ref{turb_plot}(d). In order to minimize the effect of noise distribution on the estimation of  $\Delta(\theta_s|\theta_i;z)$, we select a region of $P(\theta_s,\theta_i;z)$ as shown by the dotted red box in Fig.~\ref{turb_plot}(c).

%
%

\section{Measurement of the two-photon OAM probability distribution in turbulence}

In this section, we present a theoretical model to evaluate the influence of turbulence on the conditional OAM distribution $P(l_s|l_i; z)$. Within paraxial approximation and the Gaussian pump beam assumption \cite{kulkarni2017natcom,leach2010science}, the OAM remains conserved in SPDC. This means that if the idler photon is detected with OAM $l_i\hbar=0$, the signal photon is guaranteed to be detected with OAM $l_s\hbar=0$. Such a signal mode can be represented as: $\psi_s(\bm{\rho}'_s)=\exp\left[-\rho'^2_{s}/{4\sigma_r^2}\right]$. For evaluating the influence of turbulence on the conditional OAM distribution $P(l_s|l_i=0;z)$ of the signal photon, we simply need to evaluate how the Gaussian mode $\psi_s(\bm{\rho}'_s)=\exp\left[-\rho'^2_{s}/{4\sigma_r^2}\right]$ gets affected by turbulence. For this purpose, we calculate the cross-spectral density function of the signal photon right after the turbulence plane $z=d$ [see Fig.~\ref{turb_plot}(a)]. From $z=0$ up to $z=d$, the signal field $\psi_s(\bm{\rho}'_s)$ remains pure. Therefore, the cross-spectral density function $W(\bm{\rho}'_{s2}, \bm{\rho}'_{s1}; d)$ right after the turbulence plane $z=d$ can be written as $W(\bm{\rho}'_{s2}, \bm{\rho}'_{s1}; d) 
= \psi^*(\bm{\rho}'_{s2};d)\psi(\bm{\rho}'_{s1};d)  W_{\rm turb}(\bm{\rho}'_{s2}, \bm{\rho}'_{s1}) $, where $\bm{\rho}'_{s} \equiv (x'_{s}, y'_{s})$, is the transverse co-ordinates of signal photon at $z=d$ plane. The term $\psi^*(\bm{\rho}'_{s2};d)\psi(\bm{\rho}'_{s1};d)$ is the  cross spectral density function of the signal photon at $z=d$ right before the turbulence plane. $W_{\rm turb}(\bm{\rho}'_{s2}, \bm{\rho}'_{s1})$  is the cross-spectral density introduced by the turbulence. We approximate it as $W_{\rm turb}(\bm{\rho}'_{s2}, \bm{\rho}'_{s1})=\exp\left[-|\bm{\rho}'_{s2}-\bm{\rho}'_{s1}|^2/(2r^2)\right]$, where $r$ is the turbulence strength. Therefore, we have
\begin{align*}
W(\bm{\rho}'_{s2},\bm{\rho}'_{s1};d)  
= \exp\left[-\frac{\rho'^2_{s1}+\rho'^2_{s2}}{4\sigma_r^2}\right]  \exp\left[-\frac{|\bm{\rho}'_{s2}-\bm{\rho}'_{s1}|^2}{2r^2}\right].  \label{eqn_two_photon_turb_oam}
\end{align*}
\begin{figure}[t!]
\includegraphics[scale=1.0]{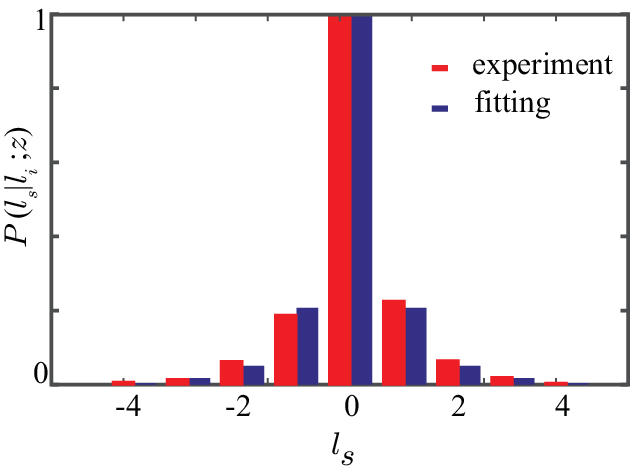}
\caption{ The experimentally measured  $P(l_s|l_i; z)$ at $z=50$ cm in turbulence. The fitting is based on the noise model described in Eq.~(\ref{oam_turb_eq}).}
	\label{turb_oam}
\end{figure}
Now, by propagating the above cross-spectral density function from $z=d$ to $z=z$, we obtain the cross-spectral density function $W(\bm{\rho}_{s1},\bm{\rho}_{s2};z)$ at $z=z$ 
\begin{multline}
	W(\bm{\rho}_{s1},\bm{\rho}_{s2};z)=\exp \left[-\frac{ik}{2R(z)}(\rho_{s2}^2-\rho_{s1}^2)\right] \\
\times \exp\left[-\frac{\rho^2_{s1}+\rho^2_{s2}}{4\sigma^2_r(z)}\right] \exp\left[-\frac{\Delta{\rho}_s^2}{2r^2(z)}\right],
\end{multline}
where $\Delta{\rho}_s=|\bm{\rho}_2-\bm{\rho}_1|$, $r(z)=r\sqrt{1+\left(\tfrac{z-z_t}{k_s\sigma_r\delta}\right)^2}$, $\sigma_r(z)=\sigma_r\sqrt{1+\left(\tfrac{z-z_t}{k_s\sigma_r\delta}\right)^2}$, $\frac{1}{\delta^2}=\frac{1}{r^2}+\frac{1}{4\sigma_r^2}$, and $k_s=\pi/\lambda_p$. We use the transformation $\bm{\rho}_{s1} \equiv (r_{s1} \cos\theta_{s1}, r_{s1} \sin\theta_{s1})$ and $\bm{\rho}_{s2} \equiv (r_{s2} \cos\theta_{s2}, r_{s2} \sin\theta_{s2})$ in order to write $W(\bm{\rho}_{s1},\bm{\rho}_{s2};z)$ as $W_s(r_s,\theta_{s1},\theta_{s2};z)$. The OAM distribution of the signal photon is same as the conditional distribution $P(l_s|l_i=0;z)$, which we write as
\begin{multline}
P(l_s|l_i=0,z)  = \iiint r_s W_s(r_s,\theta_{s1},\theta_{s2};z) e^{il_s (\theta_{s2}-\theta_{s1})} \\ \times dr_s d\theta_{s1} d\theta_{s2}.
\end{multline}
We compute the above integral numerically and find that it very closely fits the function $a\exp\left[-b|l_s|\right]$, where $a$ and $b$ are fitting parameters. We also find that $P(l_s|l_i;z)$ does not depend on $z$ after propagating through turbulence. We thus write the conditional OAM distribution as 
\begin{multline}
P(l_s|l_i,z) = a\exp\left[-b|l_s|\right]+N \exp\left[-\frac{l_s^2}{2\sigma_f^2}\right].\label{oam_turb_eq}
\end{multline}
Here, we have added the noise term, for reasons described in section~\ref{app-sec1}.  Figure~\ref{turb_oam} shows the experimentally measured $P(l_s|l_i; z)$ at $z=50$ cm. We fit  $P(l_s|l_i; z=50)$ with Eq.~(\ref{oam_turb_eq}) and obtain the experimental uncertainty $\Delta(l_s|l_i;z)$ to be $0.94\hbar$ radian$^{-1}$.

\bibliography{ref_sup}